\documentclass[twocolumn,preprintnumbers,superscriptaddress,endnote,nofootinbib,aps,prd,floatfix]{revtex4}

\usepackage{footmisc,enumerate}
\usepackage{multirow}
\usepackage{subfigure}
\usepackage{amsmath,amssymb}
\usepackage{graphicx}
\usepackage{placeins}
\usepackage{bbm,slashed}
\usepackage{xspace,slashed}
\usepackage{hyperref}
\hypersetup{colorlinks=true, citecolor=blue, urlcolor=blue, linkcolor=blue}
\usepackage[normalem]{ulem}

\usepackage[autostyle]{csquotes}

\begin{document}

\providecommand{\abs}[1]{\lvert#1\rvert}

\newcommand{\Znunujets}{(Z\to{\nu\bar{\nu}})+\text{jets}}
\newcommand{\Welnujets}{(W\to{\ell\nu})+\text{jets}}
\newcommand{\Znunujet}{(Z\to{\nu\bar{\nu}})+\text{jet}}
\newcommand{\Welnujet}{(W\to{\ell\nu})+\text{jet}}

\title{Sensing Higgs cascade decays through memory}

\begin{abstract}
Beyond the Standard Model scenarios with extensions of the Higgs sector
typically predict new resonances that can undergo a series of cascade decays to detectable Standard Model particles.
On one hand, sensitivity to such signatures will contribute to
the full reconstruction of the extended Higgs potential if a new
physics discovery will be made. On the other hand, such cascade decays
could be dominant decay channels, thus being potentially the best motivated
signatures to achieve a new physics discovery in the first place. In this work, we show 
how the long short-term memory that is encoded in the cascade decays' phenomenology
can be exploited in discriminating the signal from the background, where no such information
is present. In parallel, we demonstrate for theoretically motivated scenarios that such an
approach provides improved sensitivity compared to more standard analyses, where only information about the 
signal's final state kinematics is included.
\end{abstract}

\author{Christoph Englert} \email{christoph.englert@glasgow.ac.uk}
\affiliation{SUPA, School of Physics \& Astronomy, University of Glasgow, Glasgow G12 8QQ, UK\\[0.1cm]}
\author{Malcolm Fairbairn} \email{malcolm.fairbairn@kcl.ac.uk}
\affiliation{Theoretical Particle Physics and Cosmology, King’s College London, London WC2R 2LS, UK\\[0.1cm]}
\author{Michael Spannowsky} \email{michael.spannowsky@durham.ac.uk}
\affiliation{Institute for Particle Physics Phenomenology, Durham University, Durham DH1 3LE, UK\\[0.1cm]}
\author{Panagiotis Stylianou}\email{p.stylianou.1@research.gla.ac.uk} 
\affiliation{SUPA, School of Physics \& Astronomy, University of Glasgow, Glasgow G12 8QQ, UK\\[0.1cm]}
\author{Sreedevi Varma} \email{sreedevi.varma@kcl.ac.uk}
\affiliation{Theoretical Particle Physics and Cosmology, King’s College London, London WC2R 2LS, UK\\[0.1cm]}

\preprint{IPPP/20/36, KCL-PH-TH/2020-45}


\maketitle

\section{Introduction}
\label{sec:intro}
The search for new physics beyond the Standard Model (SM) of particle physics is
the main driver of the phenomenology programme at the Large Hadron Collider (LHC). The current negative outcome of beyond the SM (BSM) searches seems to suggest that new degrees of freedom are either too heavy or too weakly coupled to be experimentally accessible at this stage in the LHC programme. 

\begin{figure}[!t]
\includegraphics[width=8cm]{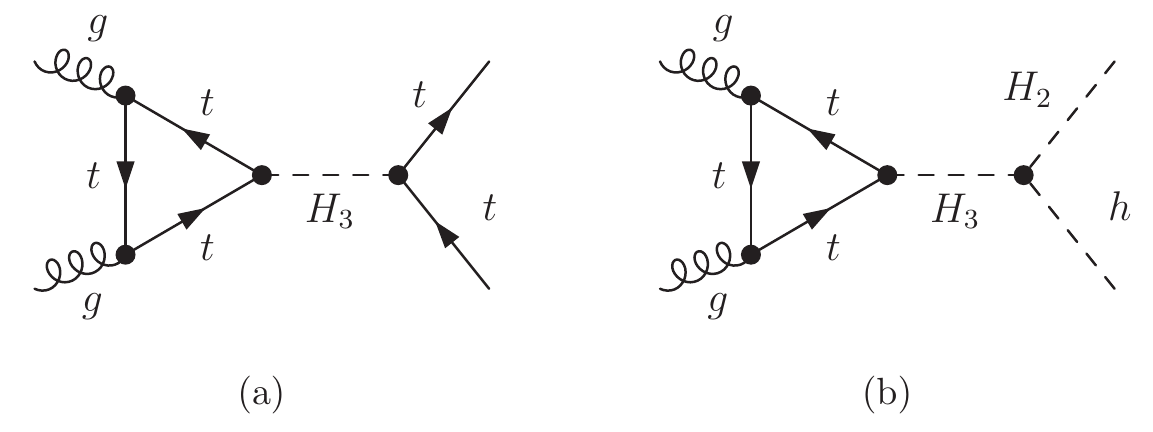}
\caption{\label{fig:gfhiggs} Representative gluon fusion diagrams for the production of an exotic scalar $H_i$ and subsequent decay into either $H_3\to t\bar t$ (a) or scalar decays $H_3\to H_2 h$ (b).}
\end{figure}

If new physics is related to the top quark and Higgs boson sector, as is expected in most concrete ultraviolet (UV) completions of the SM that tackle the shortcomings of the SM such as insufficient CP violation or TeV scale naturalness, another phenomenologically interesting avenue arises: new exotic scalar bosons could be dominantly produced through SM-Higgs like gluon fusion, Fig.~\ref{fig:gfhiggs}(a). If this production mode is relevant as a consequence of sizeable Yukawa couplings (or phases), unitarity typically implies a large decay probability into top quarks when kinematically accessible.\footnote{Decays into massive quarks are typically further enhanced due to symmetry considerations such as custodial isospin~\cite{Gunion:1989ci} or CP properties of the new scalar state~\cite{Djouadi:2005gj}.} However, it is known~\cite{Gaemers:1984sj,Dicus:1994bm,Jung:2015gta,Bernreuther:2015fts,Carena:2016npr,Djouadi:2019cbm} that large accidental interference of QCD-induced $t\bar t$ production with the scalar state can create a significant distortion of the on-shell resonance signal. When including constraints from dark matter searches, low energy experiments, flavor physics, 125 GeV Higgs signal strength measurements and exotic Higgs searches as done in Ref.~\cite{Basler:2019nas}, motivated UV completions such as the two-Higgs-doublet model (2HDM, for a review see~\cite{Branco:2011iw}) are forced into parameter regions that are particularly impacted by these interference effects.

This could mean that new physics is already present at the energy scales presently being explored by the LHC, yet interference renders the signal difficult to detect in the best motivated $t\bar t$ channel. If this is the case, sensitivity to these models can be restored using di-Higgs final states. While these final states can be enhanced by constructive signal-signal interference in concrete UV extensions of the Higgs sector~\cite{Basler:2019nas}, the significantly reduced sensitivity to such signatures will mean that new physics discoveries will be pushed into the LHC's high luminosity (HL) phase.

In scenarios with a richer scalar phenomenology, multi-Higgs production from cascade decays of a new scalar degree of freedom into a 125 GeV SM Higgs $h$ and another BSM scalar boson are possible. These signatures arise in, e.g., the next-to-minimal 2HDM~\cite{Muhlleitner:2016mzt} (N2HDM) with sizable cross sections and provide an important phenomenological input for the reconstruction of the extended Higgs potential.\footnote{See also \cite{Robens:2019kga} for a discussion of Higgs cascade decays in the context of the two-singlet extension of the SM} In scenarios like the complex 2HDM, such signatures directly probe alignment of the 125 GeV Higgs boson with fluctuations around the electroweak vacuum-independent from decoupling of additional states~\cite{Grzadkowski:2018ohf} and are therefore theoretically well motivated. Depending on the mass of the final state exotic Higgs boson, such cascade signatures also arise in the NMSSM~\cite{Basler:2018dac,Baum:2019pqc}.

In this work, we focus on decays of heavy scalars $H_3\to H_2 h$, Fig.~\ref{fig:gfhiggs} (b), where we identify the 125 GeV SM Higgs boson as the lightest scalar degree of freedom, $H_1=h$. We are specifically interested in the parameter region where $m_{H_3} > m_{H_2} + m_{h}$ and $m_{H_2}>2 m_t$, i.e. the region of parameter space where the decay 
\begin{equation}
\label{eq:branching}
H_3\to H_2 h, ~\text{with}~H_2 \to t\bar t, ~h\to b \bar b
\end{equation}
is open and sizeable. Such final states are experimentally challenging due to b-jet combinatorics and a significant amount of missing energy that renders the reconstruction of resonances difficult. While distinct kinematical correlations that are induced by the cascade decay structure can be accessed through observables like $M_{\text{T}2}$ of Refs.~\cite{Lester:1999tx,Barr:2003rg}, an analysis strategy based on rectangular combinations of collider observables might be too restrictive to obtain the highest statistical signal yield for a given background rejection.\footnote{Shower and event deconstruction are alternative all-information approaches to discriminate hadronically decaying top quarks and Higgs bosons from backgrounds \cite{Soper:2011cr, Soper:2012pb, Soper:2014rya}.} In parallel, the particular hierarchy of the branchings of Eq.~\eqref{eq:branching} induces a `timescale' for the signal events which is not present for the contributing background. While phenomenological analyses aim to perform an appropriate clustering of the final state's kinematics on a statistical level, they typically do so without accessing the event's memory imprint directly. 

Fingerprinting the relevance of this memory for signal vs background discrimination is the focus of this work. We will access this memory by means of recurrent neural networks (RNNs) and show its relevance by comparing this setup against other signal-background discrimination methodologies. 

This paper is organised as follows. In Sec.~\ref{sec:ml}, we quickly motivate the use of RNNs for the physics problem that we study in this paper and outline our analysis setup in Sec.~\ref{sec:sgnandbkg}. Sec.~\ref{sec:arch} gives an overview of the different strategies that we employ, and Sec.~\ref{sec:perf} compares the efficacy of those strategies. We summarise and conclude in Sec.~\ref{sec:conc}.

\section{Cascading memory}
\subsection{General remarks and context}
\label{sec:ml}
RNNs are networks designed to train on a sequence of time ordered events rather than spatially distributed values. In the case under consideration in this work, we employ an RNN to identify the flow of particles in the showering and branching after a collision event.
 
The architecture of the recurrent network allows it to connect a piece of information to the previous piece of information learnt, the classic example being connecting the end of the sentence to the beginning.
 
RNNs are widely used in translation (many-to-many), music generation (one-to-many) and sentiment classification or reading joined up handwriting  (many-to-one) applications. Depending on the task, recurrent neural networks might have different architectures. Our case is most similar to the structure of a many-to-one situation where the many represents the string of events as particles decay into each other and the one is the nature of the hard particles created in the initial event in the collider.
 
The RNN is a machine learning network where nodes are replaced by units, individual gates rather like logic gates but made up of algorithms consisting of fixed combinations of algebraic and smooth activation functions, as well as weights connecting those functions.  These units are then distributed across the network in the same way nodes would form a normal neural network, with freedom in the choice of architecture, e.g., the number of layers and the number of units in each layer, etc. 
 
The input is split into time ordered components like the consecutive words in a sentence and each word is fed in an ordered fashion into the first layer of units, consecutively from left to right.  Within each layer of the network,  each unit produces an activation and an output which are fed to the next layer; however, the output is also fed sideways to the right in the same layer so that the input into each unit contains information about the previous words in the sentence only.
 
In our case, the words are replaced by the parameters of jets/leptons/missing energy with the time ordering replaced by $p_T$ ordering.
 
The weights are then varied during training using the same gradient descent algorithms used for normal neural network training, and the global minimum of the cost/error function is searched for (and hopefully obtained). Updating weights requires propagation through the network of derivatives of both the cost/error function and the codependence of weights upon each other. The required chain-rule multiplication of many such derivatives increases the risk of gradients vanishing or blowing up. Hence, memorizing a longer sequence is a challenging task in traditional RNNs. Long Short Term Memory (LSTM) units can be used to construct a particular class of RNN, which address these issues and remembers information for a longer period~\cite{10.1162/neco.1997.9.8.1735}.  LSTM and closely related Gated Recurrent Unit (GRU)~\cite{cho-etal-2014-learning} networks have a gated structure that regulates the passage of information through the unit. The LSTM architecture therefore stabilises the way that the units change their behaviour as weights are updated.

Applications of RNNs to jet physics have emerged in recent years. LSTMs have been in use for flavour tagging~\cite{Guest:2016iqz, ATLAS:2017gpy}, substructure studies~\cite{Egan:2017ojy,Fraser:2018ieu,Kasieczka:2019dbj}, hardware analysis \cite{Wielgosz:2016xhl,Wielgosz:2017wtx} and event-level classifiers~\cite{Louppe:2017ipp}. Various deep learning techniques to classify light-flavoured and heavy-flavoured jets are compared in~\cite{Guest:2016iqz}. Particle tracks and vertices are used as the classifying features of the network. A comparison study comparing RNNs to deep neural networks (DNNs), LSTMs and outer recursive networks, while exploiting their prowess in tracing the full event history, was performed in this paper. The classification of jets (up quark initiated vs down quark initiated) using their electric charge was attempted in~\cite{Fraser:2018ieu}. Convolutional, recurrent, and recursive neural networks were used to train the network. 
The approach followed in~\cite{Louppe:2017ipp} is to investigate the analogy between the way RNNs perform natural language processing to training on jet physics. Jets derived from sequential clustering algorithms are fed into the network and used for classification purpose.

\subsection{Event data and pre-processing}
\label{sec:sgnandbkg}
In this work, we consider on the cascade decay signature
\begin{equation}
\label{eq:sig}
p p \to H_3 \to (H_2 \to t\bar t  \to \ell^+\ell'^- b\bar b + {\slashed{E}}_T ) + (h \to  b\bar b)\,,
\end{equation}
i.e. we feed in the two leptons, four b-jets and the missing energy as the discriminating features of the signal. Pseudorapidity $\eta$, azimuthal angle $\phi$, transverse momentum $p_T$, and energy $E$ parameters are used to pass the information into the network. We focus on 13~TeV collisions.
The signature of two leptons, four b-tagged jets, and missing energy arises when the tops decay to b quarks, leptons, and neutrinos, thus providing a range of correlations and a cluster history that is not (fully) present for the contributing background processes, which include 
\begin{itemize}
	\item $pp \to t \bar{t} b \bar{b}$, 
	\item $pp \to t \bar{t} (Z\to b \bar{b})$, 
	\item $pp \to t \bar{t} (h\to b \bar{b})$, 
	\item $pp \to b b \bar{b} \bar{b} W^+ W^-$,
	\item $pp \to b b \bar{b} \bar{b} Z Z$.
\end{itemize}
Out of these, the $t\bar t b\bar b$ production is by far the most dominant contribution; see Tab.~\ref{tab:numevents}. 

\begin{table}[!t]
	\centering
	\begin{tabular}{lc}
	\hline
	 Process                               &   Cross Section [fb] \\
	\hline                                                          
	 $pp \to t \bar{t} b \bar{b}$          &             1215.050 \\
	 $pp \to t \bar{t} (h\to b \bar{b})$   &               22.007 \\
	 $pp \to t \bar{t} (Z\to b \bar{b})$   &                6.096 \\
	 $pp \to b b \bar{b} \bar{b} W^+ W^-$  &                2.561 \\
	 $pp \to b b \bar{b} \bar{b} Z Z$      &                0.014 \\
	\hline
	\end{tabular}	
	\caption{Inclusive cross sections for background processes at a 13 TeV LHC, where $t \bar{t} b \bar{b}$, $t \bar{t} h$ and $t \bar{t} Z$ normalisations include K-factors $1.8$~\cite{Bredenstein:2009aj}, $1.17$~\cite{deFlorian:2016spz}, and $1.2$~\cite{Atlas:2019qfx}, respectively.\label{tab:numevents}}
\end{table}

The signal is modelled with {\sc{FeynRules}}~\cite{Christensen:2008py,Alloul:2013bka} and we generate signal and background events with {\sc{MadEvent}}~\cite{Alwall:2011uj,deAquino:2011ub,Alwall:2014hca}
and {\sc{MadSpin}}~\cite{Frixione:2007zp,Artoisenet:2012st}. The generated events are showered with {\sc{Pythia8}}~\cite{Sjostrand:2014zea} and outputted in the {\sc{HepMC}} format~\cite{Dobbs:2001ck}. We use {\sc{FastJet}}~\cite{Cacciari:2011ma,Cacciari:2005hq} for clustering jets, interfaced through the reconstruction mode of {\sc{MadAnalysis}}~\cite{Conte:2012fm,Conte:2014zja,Dumont:2014tja,Conte:2018vmg}. All jets are clustered with the anti-kT algorithm~\cite{Cacciari:2008gp} of radius $0.4$ with the requirement that they have a transverse momentum of 
\begin{equation}
\label{eq:cuta}
p_T(j)  >  20~\text{GeV}
\end{equation}
and a pseudorapidity of 
\begin{equation} 
\abs{\eta(j)} < 4.5\,.
\end{equation}
B-jets are selected with efficiency of $\epsilon=0.8$ and in the central part of the detector within 
\begin{equation}
\label{eq:btag}
\abs{\eta(j_b)} < 2.5\,.
\end{equation}
The final state leptons are selected if  
\begin{equation}
\label{eq:cutb}
p_T (\ell) > 5~\text{GeV}~\text{and~}\abs{\eta(\ell)} < 2.5\,.
\end{equation} 
Subsequently, we impose isolation criteria, where a lepton is considered isolated if the total $p_T$ of the jets within the light lepton's cone radius $R = \sqrt{(\Delta \eta )^2 + (\Delta \phi )^2} = 0.3$ is less than $20 \%$ of the lepton's transverse momentum $p_T (\ell)$. The event is accepted if exactly two leptons and four b-tagged jets are identified, otherwise vetoed. The missing transverse momentum is evaluated as the opposite to the four-momenta sum of jet and lepton tracks in the plane perpendicular to the beam and its magnitude is considered as the missing transverse energy. 

\begin{figure*}[!t]
\begin{center}
	\subfigure{\includegraphics[width=0.48\textwidth]{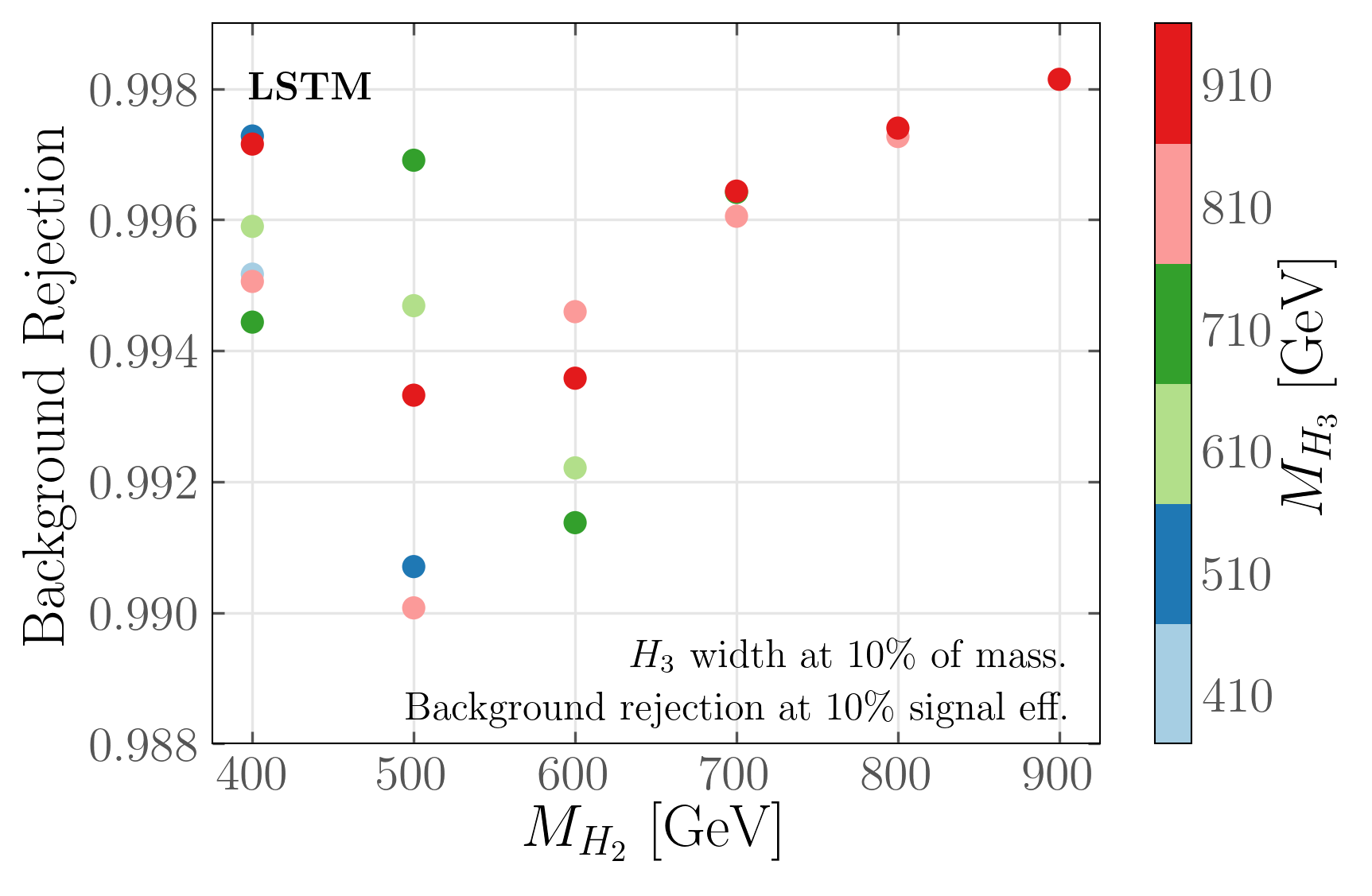}}
   \hskip 0.5cm
   \subfigure{\includegraphics[width=0.48\textwidth]{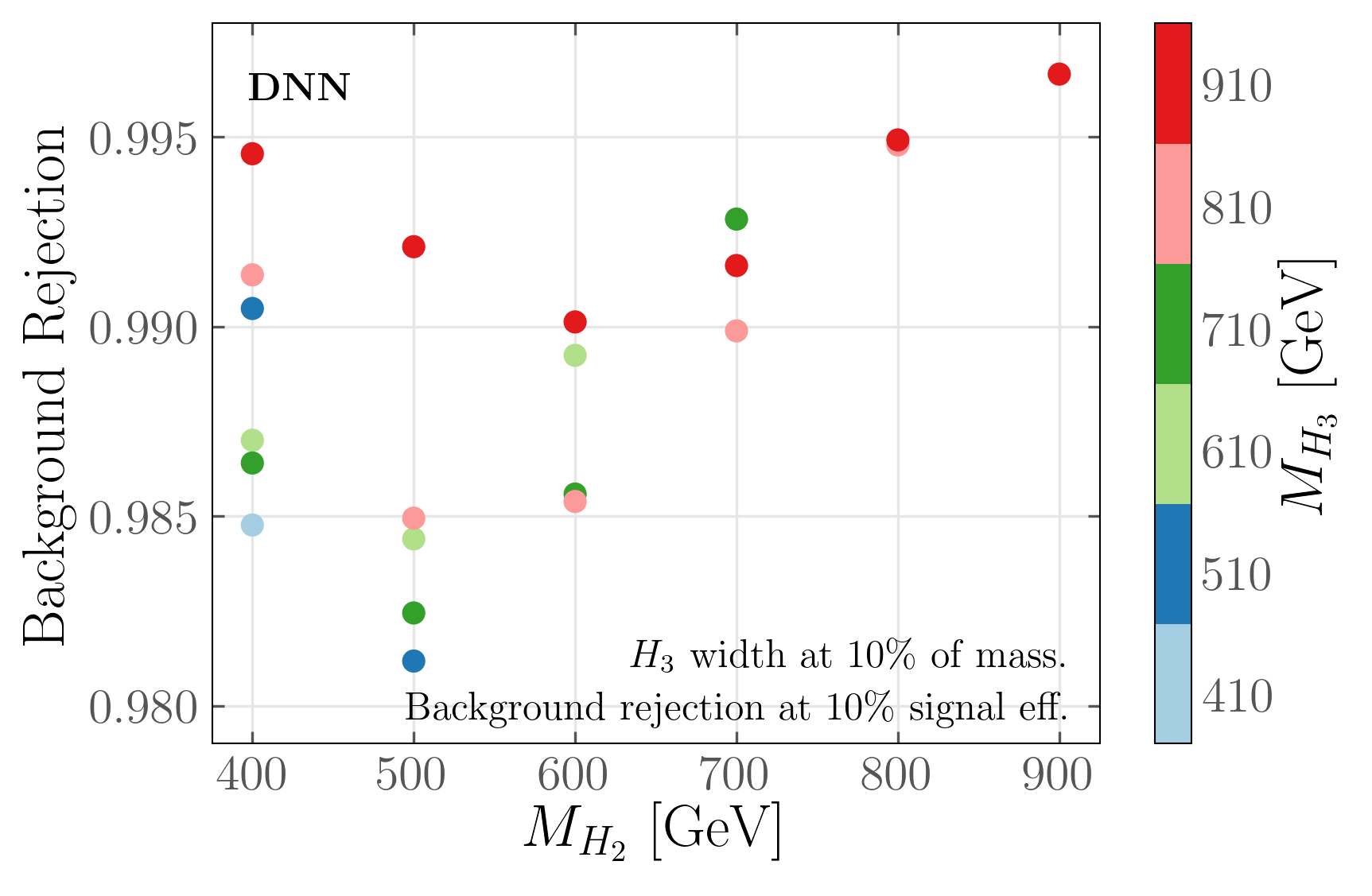}}
   \vskip\baselineskip
   \subfigure{\includegraphics[width=0.48\textwidth]{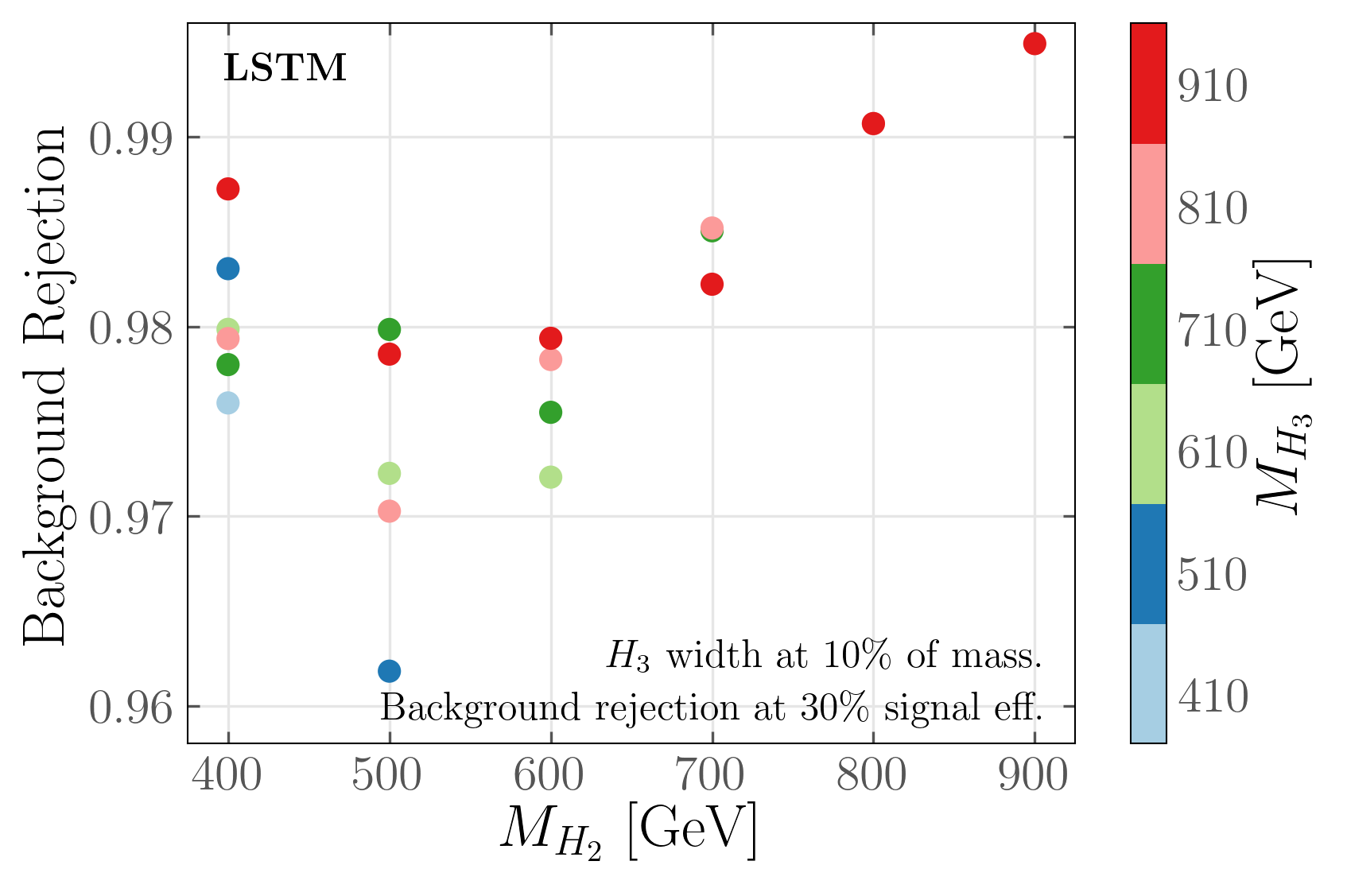}}
   \hskip 0.5cm
   \subfigure{\includegraphics[width=0.48\textwidth]{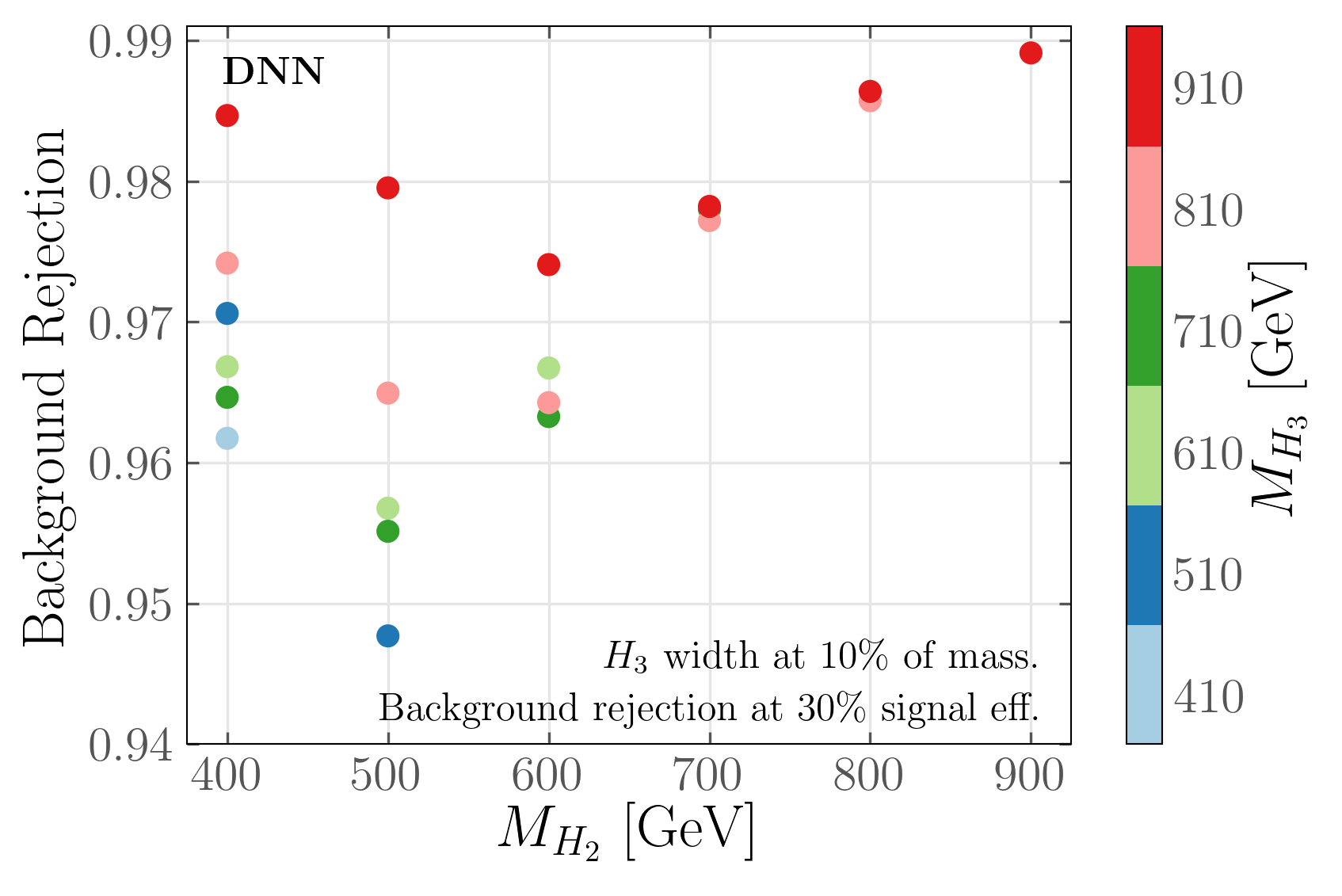}}
	\caption{Figures showing the background rejection of LSTM and DNN networks for signal efficiencies of $10\%$ and $30\%$. The signal is generated with different masses for each case and $M_{H_3} > M_{H_2}+m_h$. The width of ${H_3}$ is set to $10\%$ of its mass, while the width of $H_2$ is calculated assuming 100\% branching into top quarks. Note that these assumptions predominantly influence the normalisation and not the efficiency of the networks. For each signal case, the network is trained along with the background events for a luminosity of $500/$fb. The LSTM network has one LSTM layer of $45$ units, a dropout rate of $0.1$ and learning rate of $0.001$, while the DNN network has two fully connected layers of $80$ units and a learning rate of $0.001$. The mass scans are performed using Monte Carlo truth particles before showering and particle reconstruction. The networks use $10000$ events split into training, testing and validation sets. Runs with the width of $H_3$ set to $30\%$ of its mass produced comparable results.\label{fig:mass_scan}}
\end{center}
\end{figure*}

The production of $t \bar{t} b \bar{b}$ provides the largest background contribution. $b b \bar{b} \bar{b} Z$ is many orders of magnitude smaller, and we will not consider it further. Event numbers are rescaled to a sum of $69000$ events before passed to the neural network with an equal number of signal events for training. 

\subsection{Architectures}
\label{sec:arch}
Before we turn to an application of the RNN strategy to analyse an actual N2HDM scenario, we would like to quantify the information gain that becomes available by using RNN as opposed to other strategies that do not directly access the memory of the signal event decay chain. To this end, the data were trained using two different networks to assess the importance of long and short term memory on the classification score, i.e., we compare the performance of the RNN with that of a DNN. Models are built using Tensorflow2.1 \cite{abadi2016tensorflow} and trained using \textsc{Nvidia GeForce GTX 1080Ti} and \textsc{RTX 2080Ti} on the \textsc{CUDA} 10.2 platform~\cite{Nickolls:2008:SPP:1365490.1365500}. We use 81\% of events for training, 9\% for validation, and 10\% for testing.

The RNN network is trained for many different architectures - we vary the number of GRU (LSTM) layers from one to nine, while the number of RNN units also vary from 10 to 100. Default parameters are used in the GRU/LSTM units,  and Tanh activation is applied to the units while the sigmoid function is used as the recurrent activation. Weights are initialized using Glorot~\cite{pmlr-v9-glorot10a} uniform initializers in the GRU/LSTM units while orthogonal initializers are used in the recurrent states. A dropout parameter of 0.1 is used between the layers to avoid overfitting. 

The DNN is trained  using identical hyperparameters (without dropout). The number of layers and the number of units are varied as in the RNN. Weights of the DNN are also initialized using Glorot~\cite{pmlr-v9-glorot10a} and ReLU ~\cite{Nair:2010:RLU:3104322.3104425} activation was given to the layers.

Both the RNN and the DNN are optimized using the Adam~\cite{DBLP:journals/corr/KingmaB14} algorithm using categorical cross-entropy with a learning rate of 0.001 and default beta parameters, which are the exponential decay rates for the first- (0.9) and second-moment estimates (0.99), respectively. The networks are trained for 100 epochs with early patience of 10. The output layer is activated using softmax~\cite{10.1007/978-3-642-76153-9_28} activation to obtain the class probability (binary cross-entropy with the output layer activated using a sigmoid function produces similar results).

\subsection{Performance comparison and physics}
\label{sec:perf}

\begin{figure}[!t]
\begin{center}
   \subfigure{\includegraphics[width=0.45\textwidth]{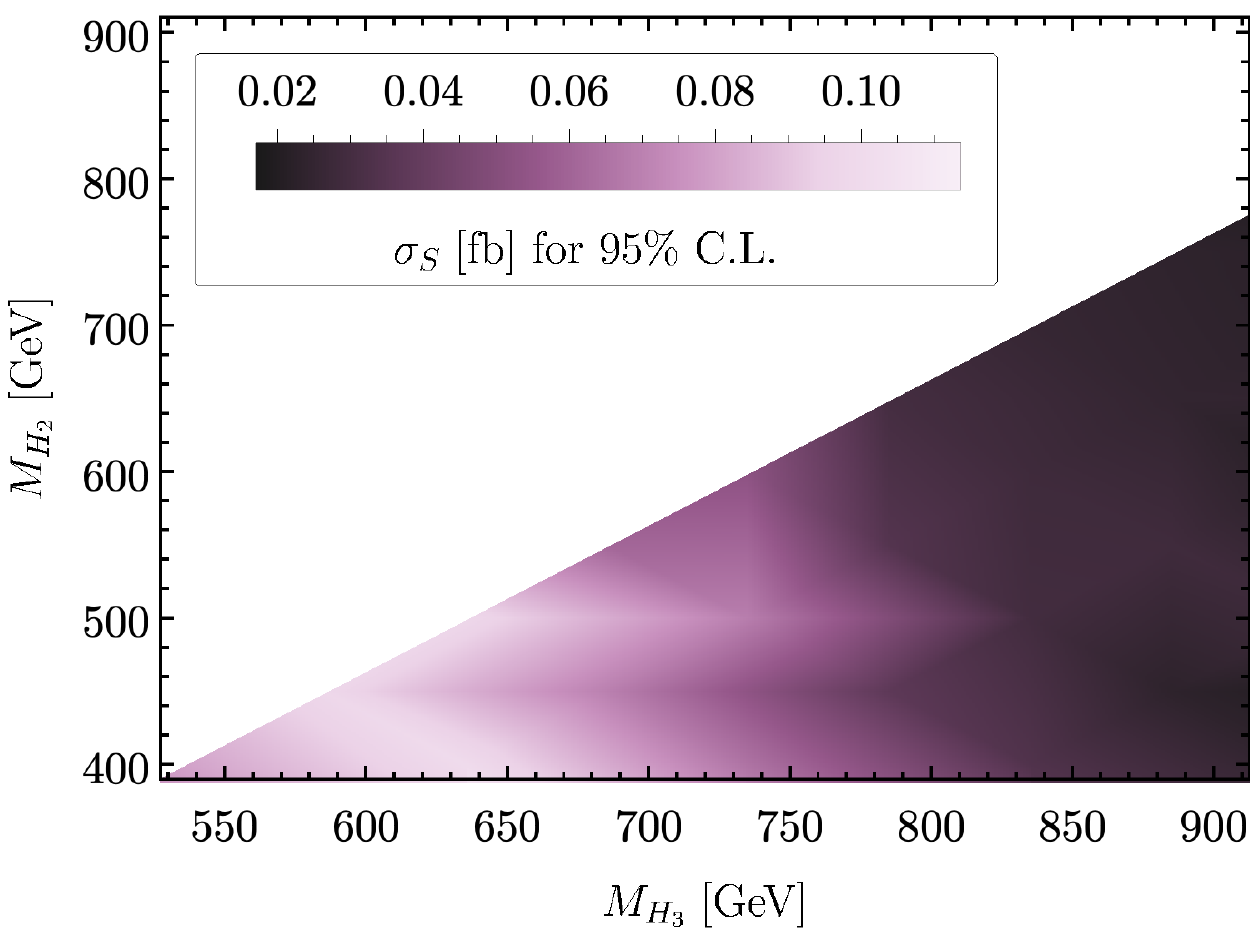}}
	\caption{Sensitivity in the $M_{H_3}-M_{H_2}$ plane displayed as the final signal cross section $\sigma_S$ required to achieve $S/\sqrt{B} = 2$ ($95\%$~C.L.), using the LSTM framework described in the text. We scan over the mass parameters with the requirement $M_{H_3} > M_{H_2} + 125$~GeV and fixed branching ratios $\text{BR}(H_3 \to H_2 h) = 0.5$ and $\text{BR}(H_2 \to t \bar{t}) = 1$.  \label{fig:xsecsens}}
\end{center}
\end{figure}

To check whether an LSTM/GRU network provides additional sensitivity in events with long decay chains, we perform a scan over signal configurations with masses $M_{H_3}$, $M_{H_2} \in [410, 950]$~GeV and $H_3$ width at $10\%$ and $30\%$ of $M_{H_3}$. LSTM and DNN networks are trained on each signal sample evaluating the background rejection of the network when minimum signal efficiencies of $10\%$ and $30\%$ are required. The model-independent scan over masses considers directly the MC-truth information without showering or realistic selection criteria to clarify the a priori usability of both types of neural networks for the considered cascade decays.
The LSTM network shows an improved performance compared to the DNN, especially when larger signal efficiencies are required as shown in Fig.~\ref{fig:mass_scan}. The kinematical observables of the final state particles largely depend on the mass of the different resonant structures in the BSM. Therefore, the networks are able to better discriminate from the SM background for larger $H_2$, $H_3$ masses that lead to a more pronounced cascade decay phenomenology. This leads to the reduction of the background and opens up the possibility of excluding points of the parameter space of concrete scenarios. Including effects from showering and hadronisation (which creates additional sources of missing energy from meson decays), and realistic acceptance criteria Eqs.~\eqref{eq:cuta}-\eqref{eq:cutb}, we show sensitivity projections for the HL-LHC (3/ab) as a function of $M_{H_{3,2}}$ in Fig.~\ref{fig:xsecsens}. As the background rapidly falls with $M_{H_3}$ mass hypothesis we are sensitive to smaller cross sections at higher mass. 

\begin{figure*}[!t]
\begin{center}
	\subfigure[\label{fig:loss}]{\includegraphics[width=0.48\textwidth]{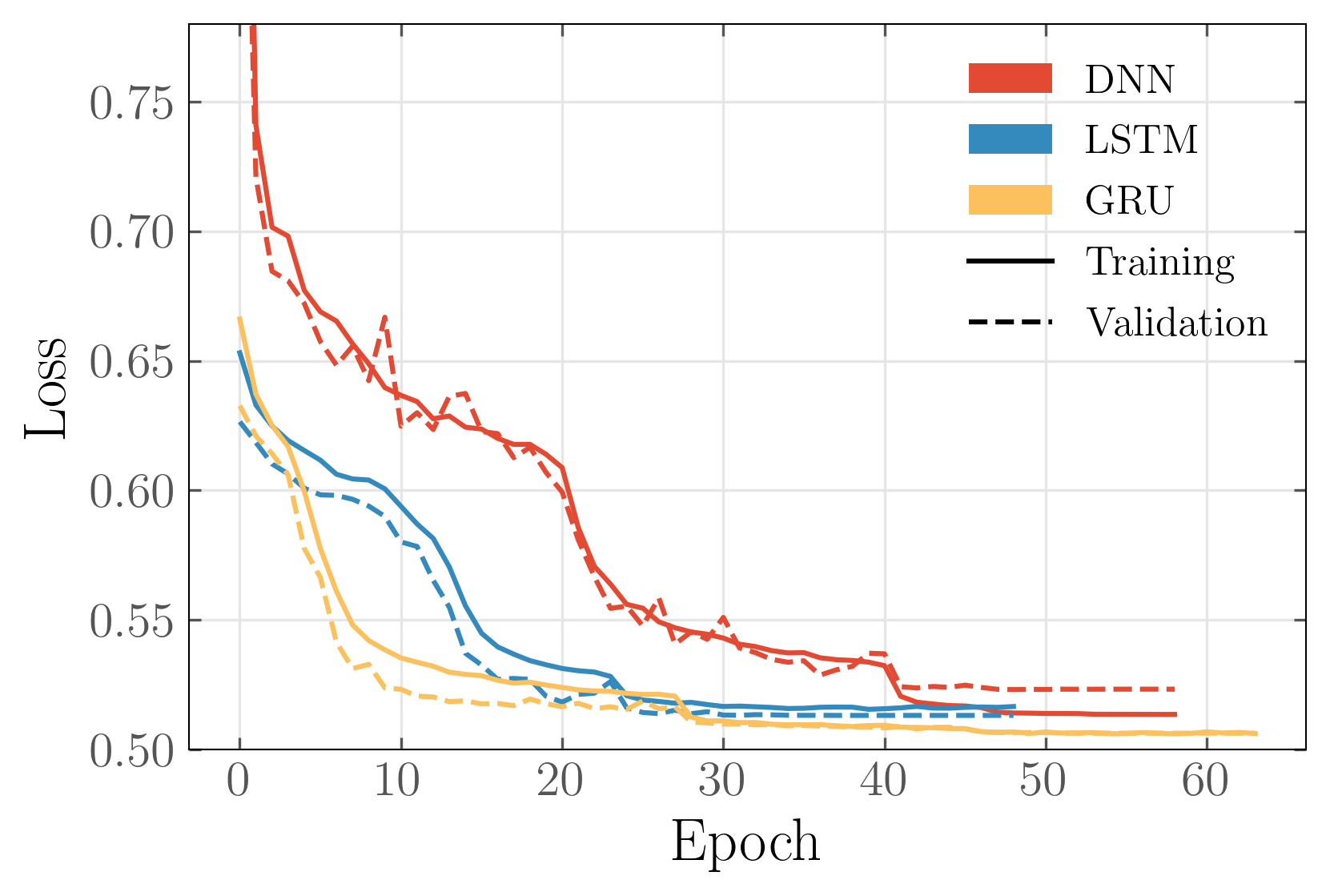}}
   \hskip 0.5cm
	\subfigure[\label{fig:roc}]{\includegraphics[width=0.48\textwidth]{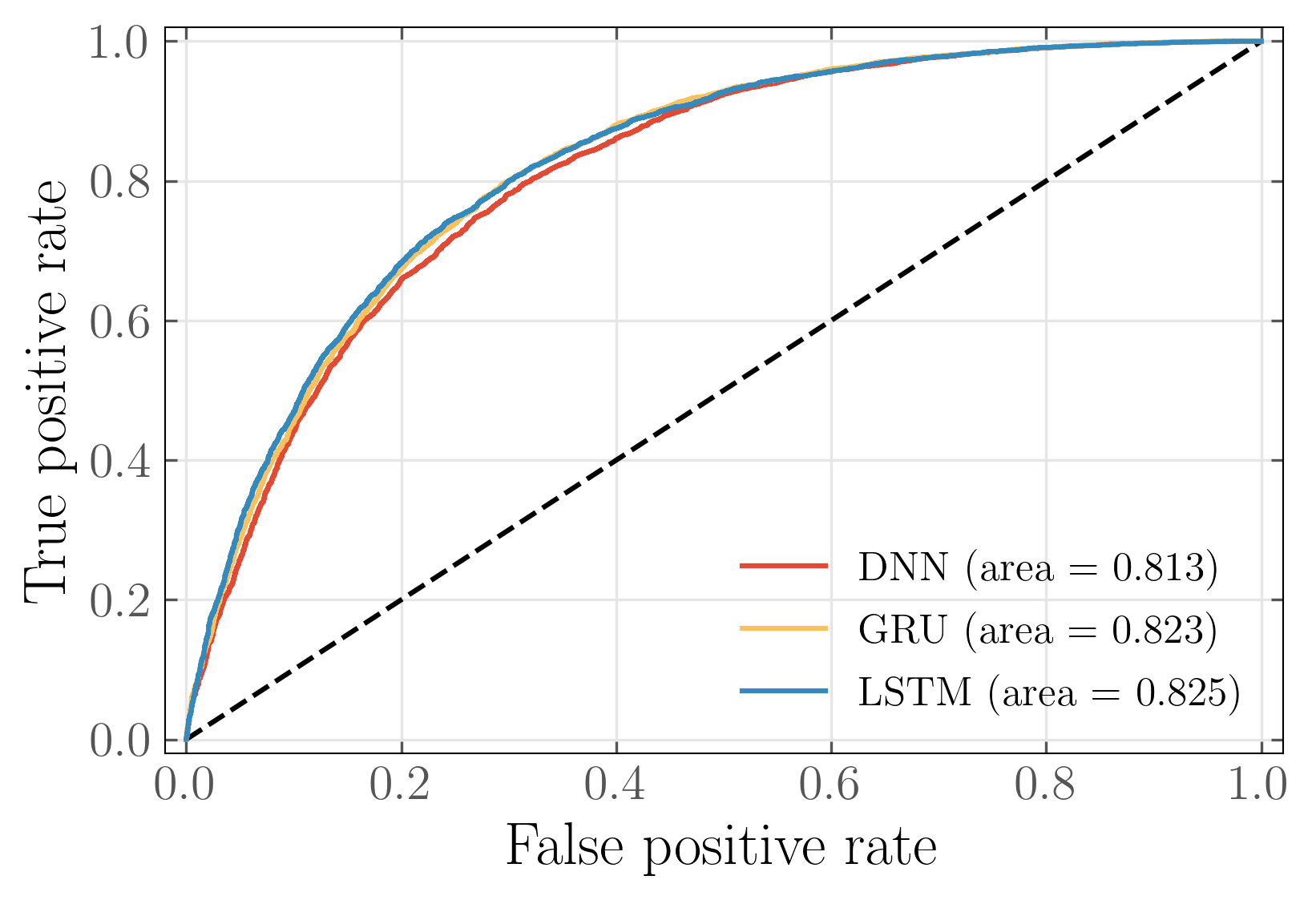}}
   \vskip\baselineskip
	\subfigure[\label{fig:rnnscore}]{\includegraphics[width=0.48\textwidth]{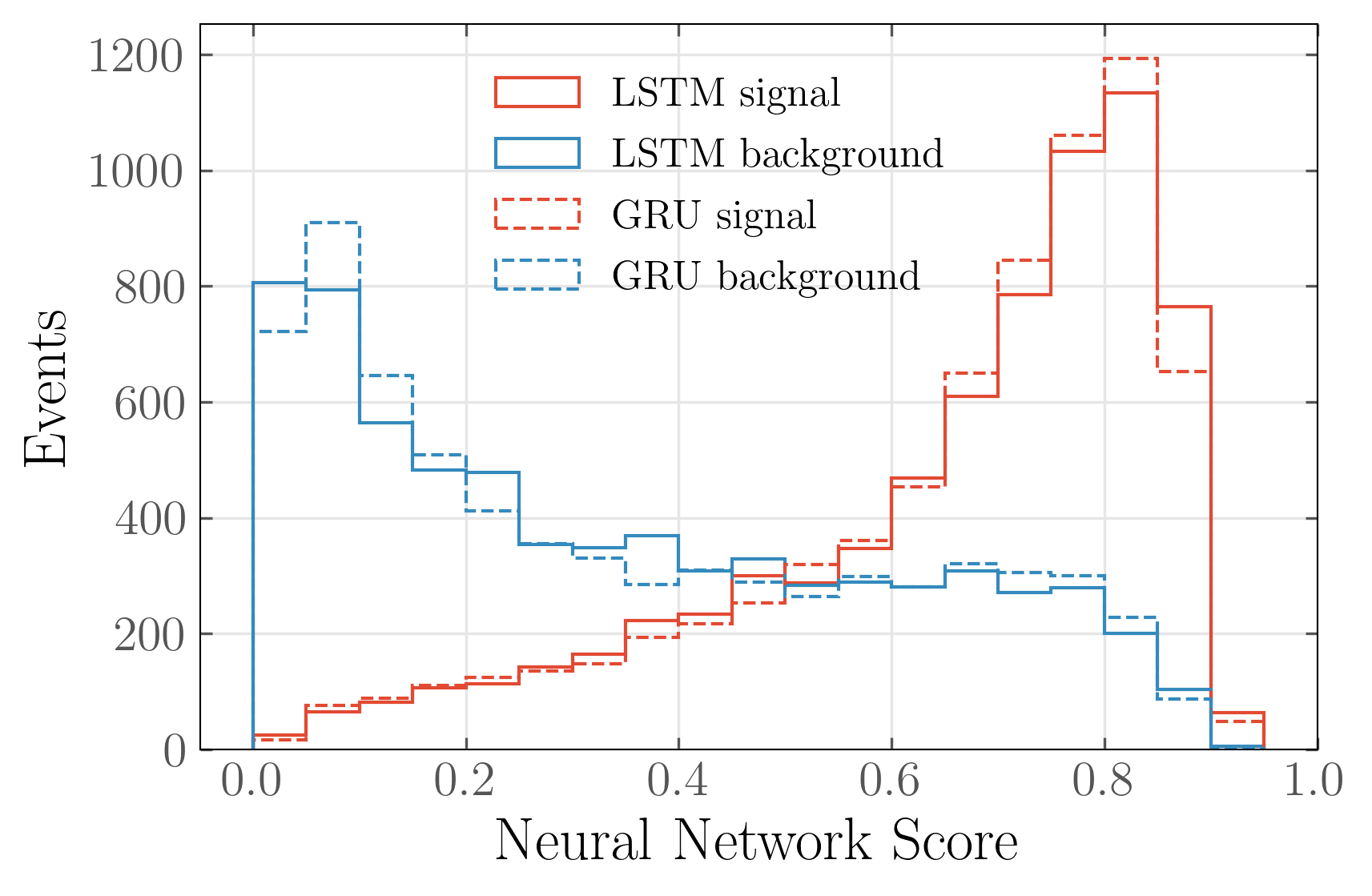}}
   \hskip 0.5cm
	\subfigure[\label{fig:dnnscore}]{\includegraphics[width=0.48\textwidth]{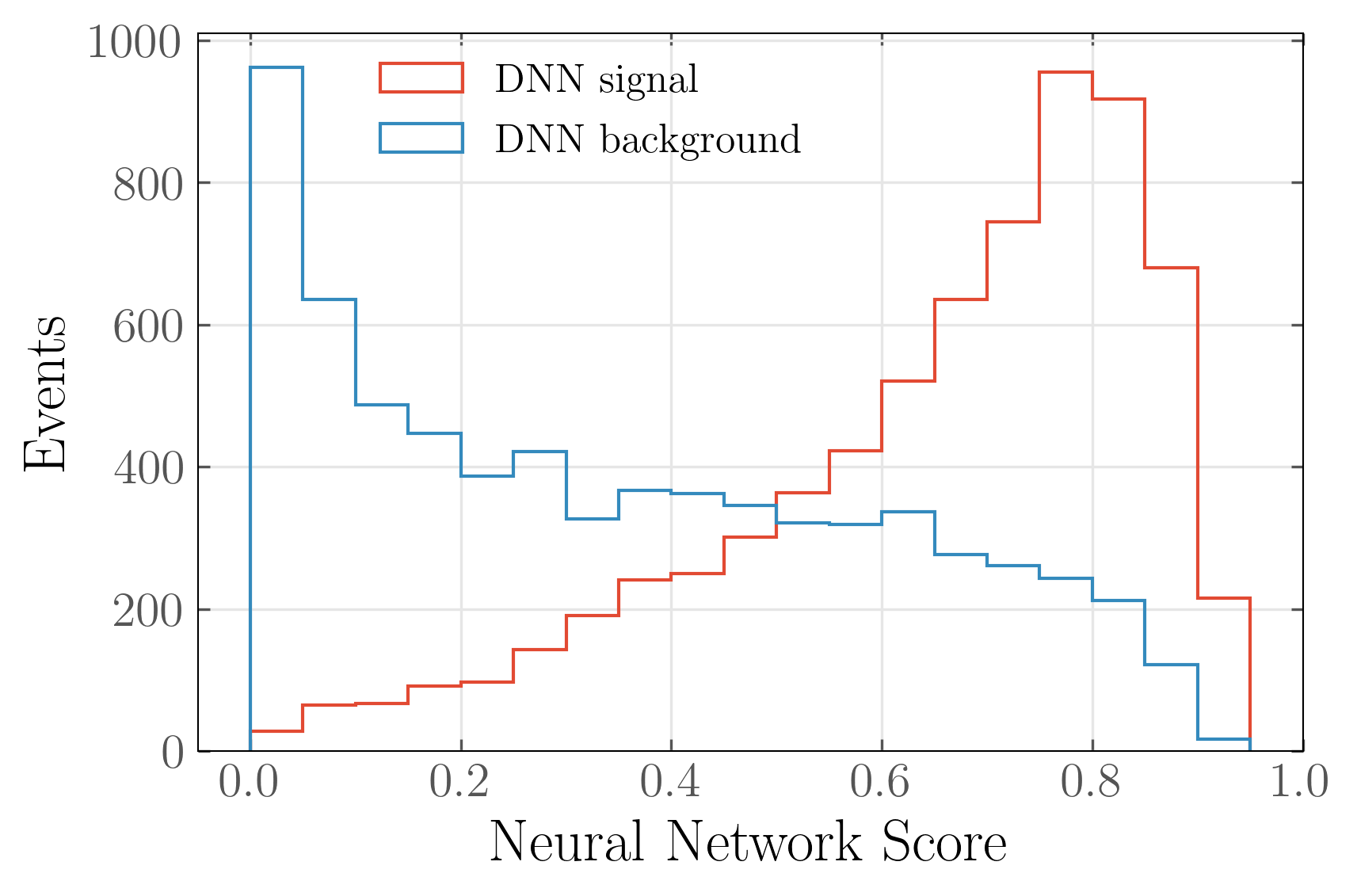}}
	\caption{\label{fig:ML} (a) Loss curves for training and validation data for LSTM, GRU and DNN along with (b) ROC curves for each case for the N2HDM benchmark masses (see text). Class probability values for the (c) RNN cases and for (d) DNN. The LSTM (GRU) consists of one layer of 45 units which result in comparable performance while the DNN is built with two dense layers of 80 units. The latter provides slightly poorer discrimination of signal against background and requires more epochs to minimise the loss function.
	}
\end{center}
\end{figure*}

\begin{figure*}[!t]
\begin{center}
   \subfigure{\includegraphics[width=0.45\textwidth]{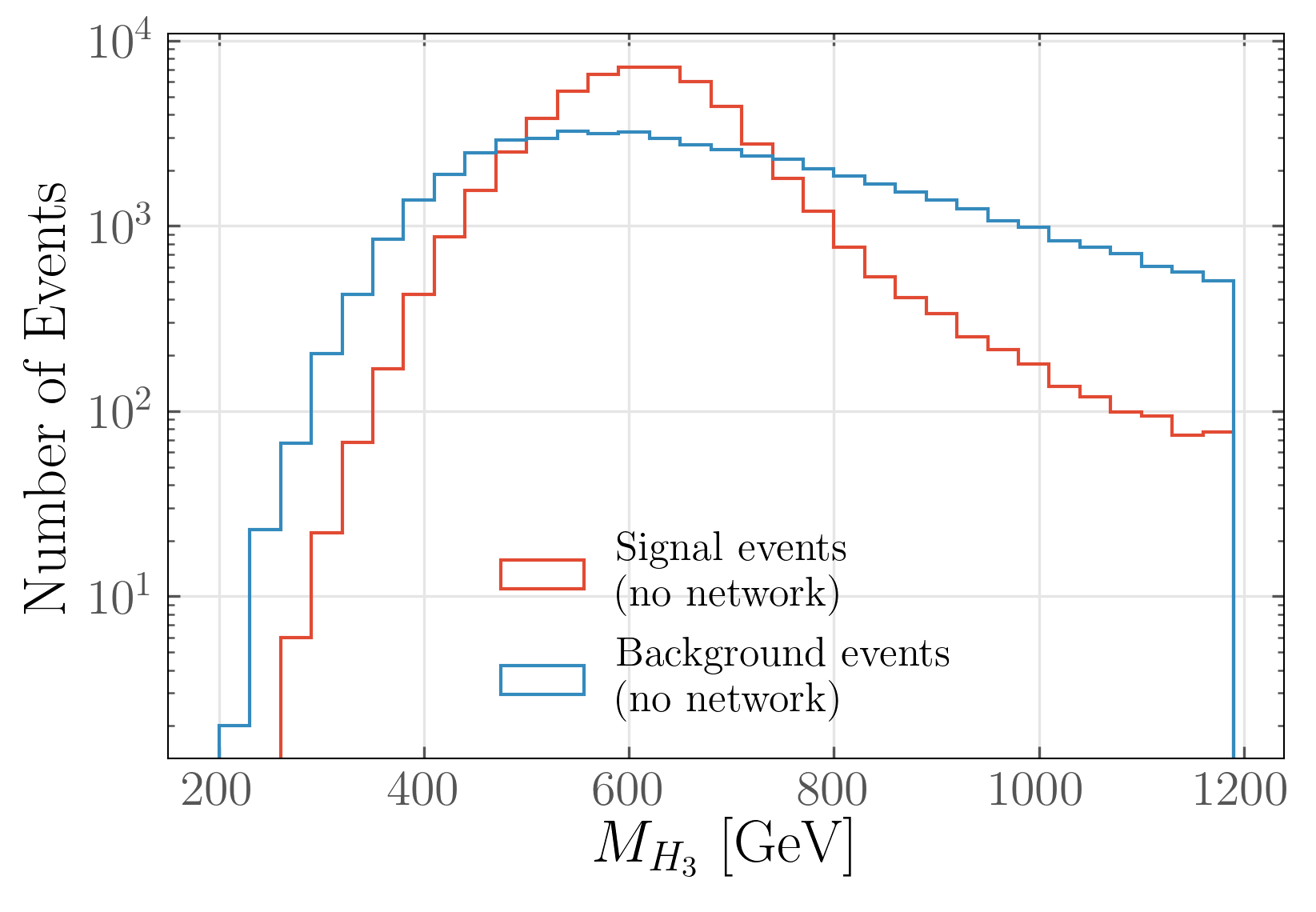}}
   \hskip 0.5cm
   \subfigure{\includegraphics[width=0.45\textwidth]{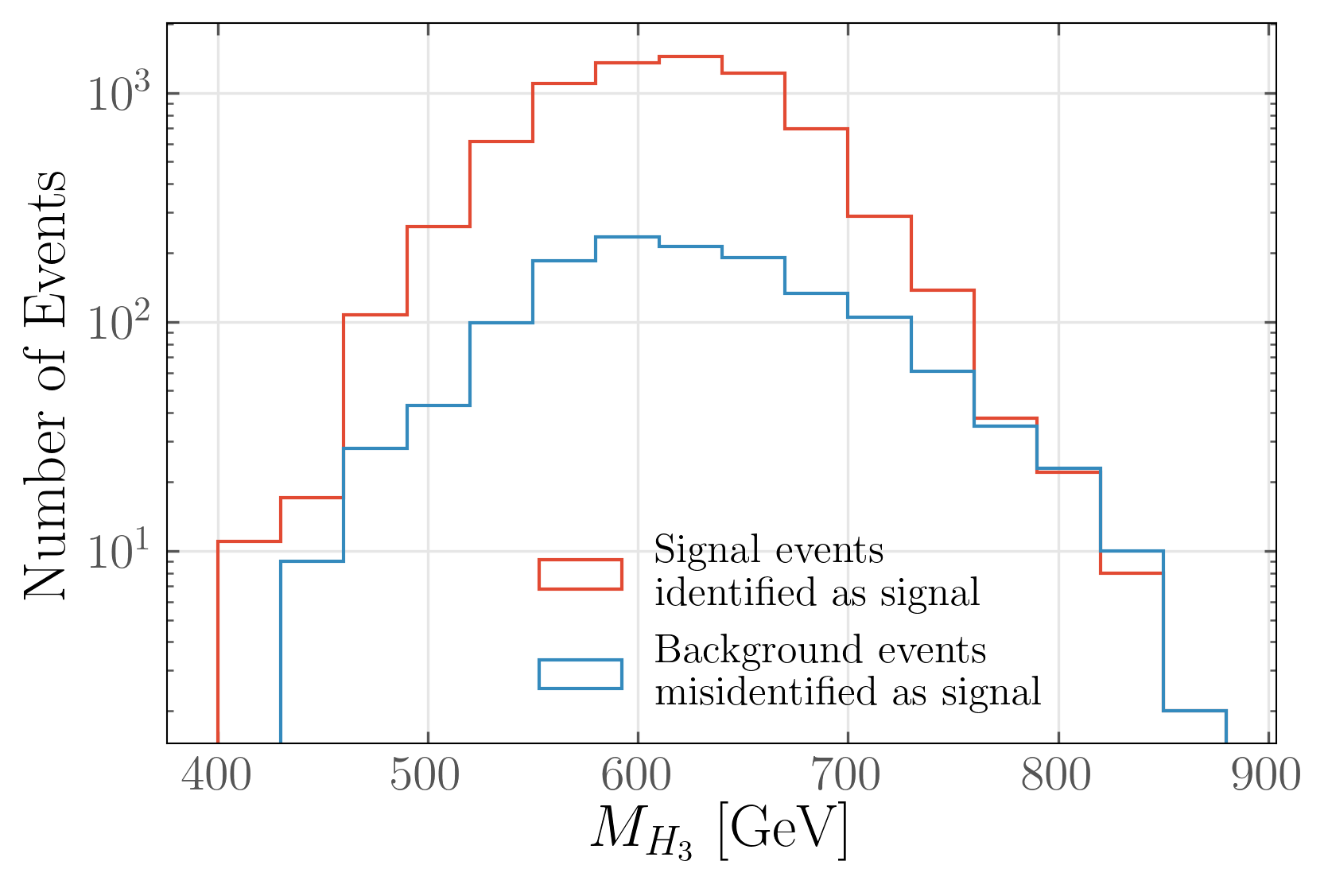}}
   \vskip\baselineskip
   \subfigure{\includegraphics[width=0.45\textwidth]{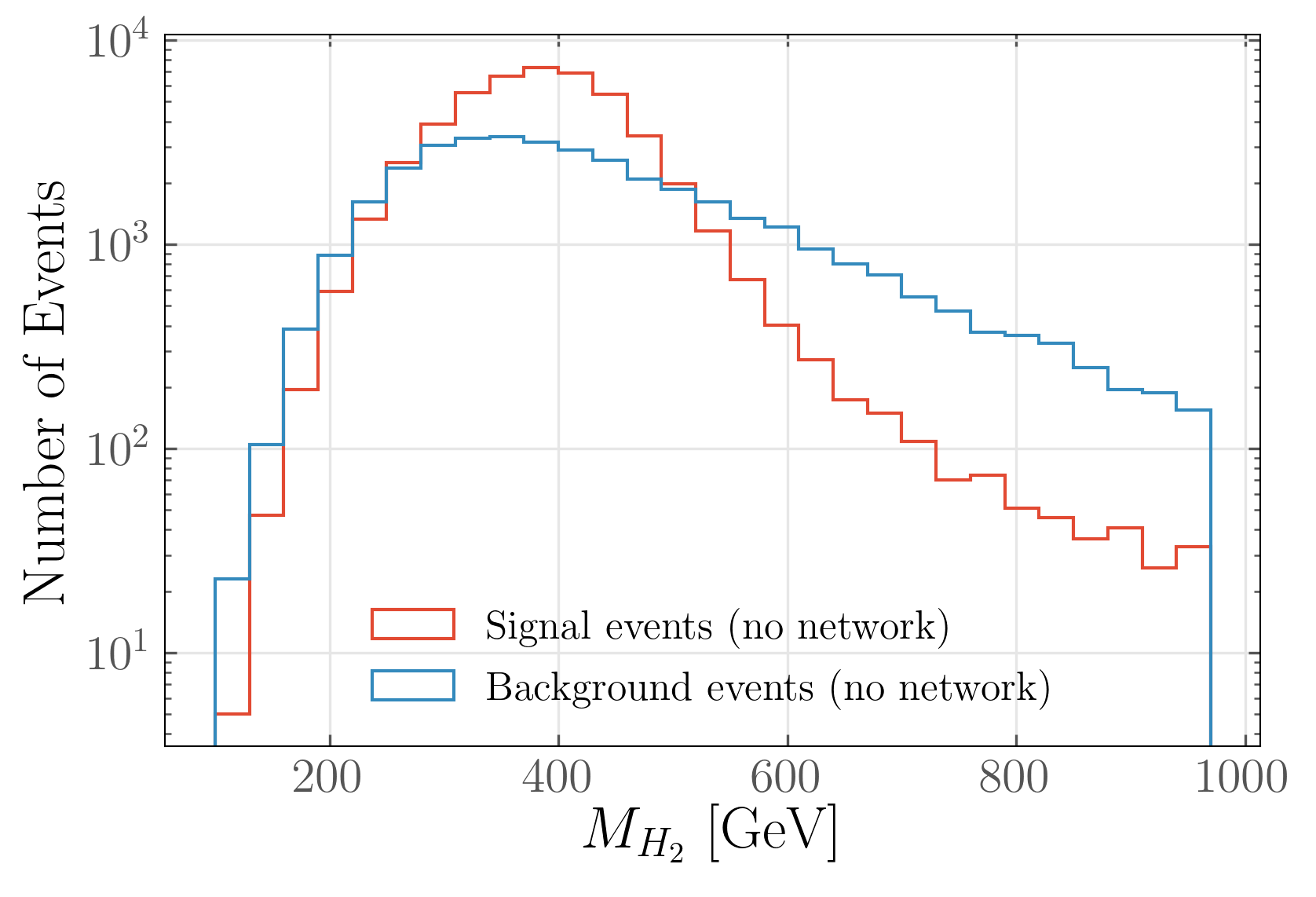}}
   \hskip 0.5cm
   \subfigure{\includegraphics[width=0.45\textwidth]{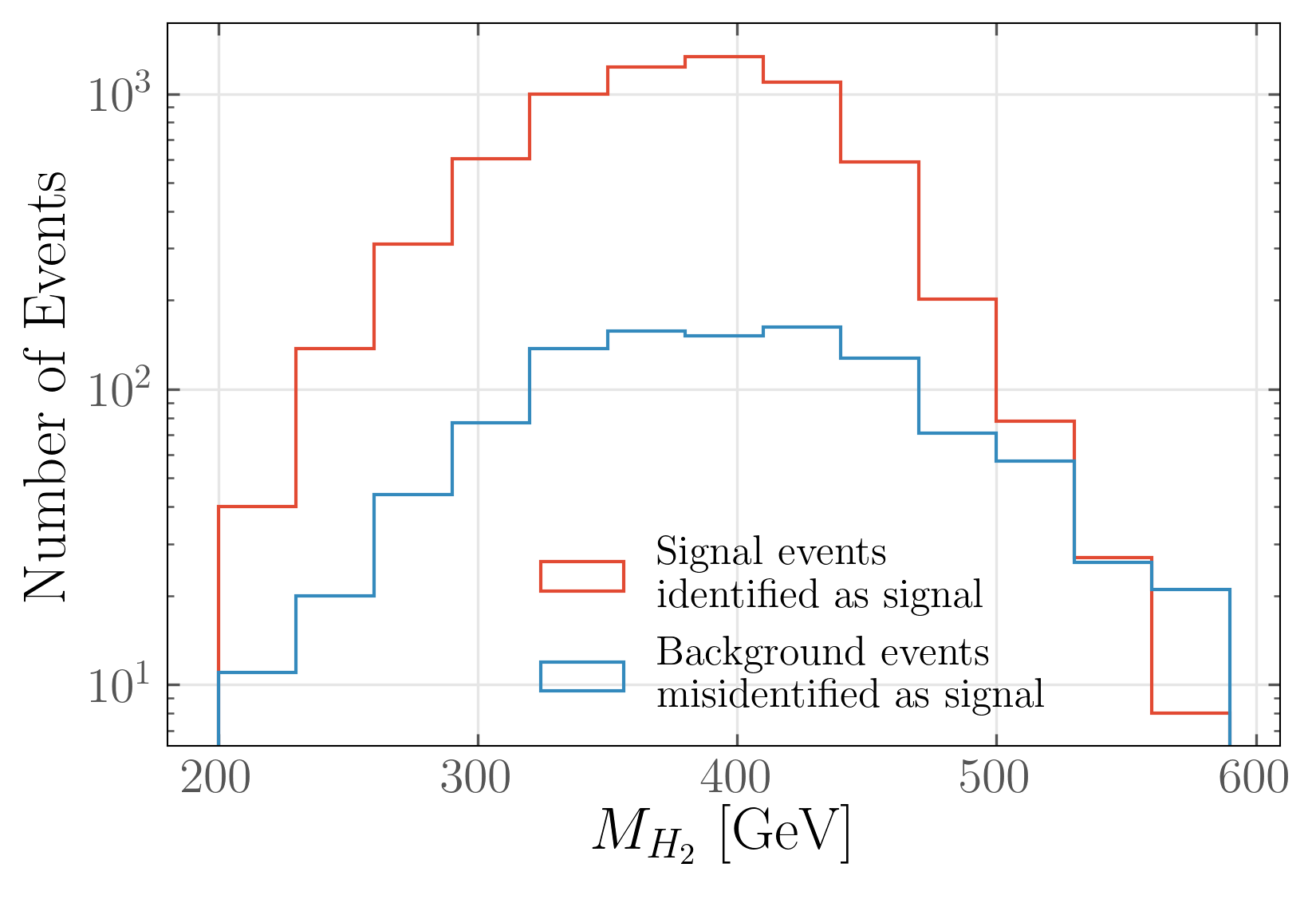}}
    \caption{Example histograms with and without the LSTM neural network for an N2HDM set of couplings with $M_{H_2} = 480$~GeV, $M_{H_3} = 722$~GeV, and widths $4.9$ and $45$~GeV for $H_2$ and $H_3$, respectively. The LSTM network used had one LSTM layer of 45 units, a dropout rate of 0.1, and learning rate of 0.001. \label{fig:reso_histo}}
\end{center}
\end{figure*}

Given the RNN network's enhanced sensitivity to the cascade decay's phenomenology, we can further discuss its relevance for motivated scenarios beyond the generic scan of Fig.~\ref{fig:xsecsens}. To this end, we focus on the N2HDM as a prototype scenario that predicts the signature of Eq.~\eqref{eq:sig}. Relevant coupling points are obtained by scanning the parameter space of the N2HDM using {\sc{ScannerS}}~\cite{Coimbra:2013qq,Ferreira:2014dya,Costa:2015llh,Muhlleitner:2016mzt,Muhlleitner:2020wwk} and requiring the branching ratios of the scalars $\text{BR}( H_3 \to H_2 h )$, $\text{BR}( H_2 \to t \bar{t})$, and the pseudoscalar $\text{BR}( A \to t \bar{t} )$ to be larger than $0.5$.\footnote{CMS and ATLAS are searching for charged~\cite{Aaboud:2018cwk,Sirunyan:2020hwv} and neutral Higgs bosons~\cite{Sirunyan:2019wrn,CMS:2019lei,Aaboud:2017hnm}, which can provide additional sensitivity to this parameter region.} To demonstrate the sensitivity that is available through the GRU/LSTM setup, we focus on a N2HDM parameter point with $M_{H_2} = 480$~GeV, $M_{H_3} = 722$~GeV, and widths $4.9$ and $45$~GeV for $H_2$ and $H_3$, respectively. This point has a cross section of $3.43$~fb and passes the branching requirements with $\text{BR}(H_3\to H_2 h) = 0.52$. QCD corrections for the signal are included via reweighting to the Higgs Cross Section Working Group values~\cite{Dittmaier:2011ti} (see also \cite{Dawson:1990zj,Graudenz:1992pv}) for the scale choice of $\mu=M_{H_3}/2$. Again, we include the effects of showering as well as additional sources of missing energy that arise from hadronisation and meson decays.

As usual in machine learning, the performance of the classifiers is visualised using the Receiver Operating Characteristics (ROC) curves. Histograms of the class probability values are also plotted to check how well the predicted probability values are separated for a given classification threshold. 
Again, we train the networks with different hyperparameters
and ROC curves of networks with good performance on both training and validation data, which are shown in Fig.~\ref{fig:ML}.
We find that the RNN always shows a slightly better discrimination.
This leads us to the conclusion that the splitting history encoded in the event indeed provides relevant information that allows one to discriminate the background from the signal in a slightly more nuanced way.  While the cascade decay leaves discriminating features in the final state kinematic information which are used by the DNN to perform the classification, the quicker convergence of the RNN setup demonstrates that this information is more efficiently learned through an adapted architecture that reflects the branching hierarchy directly.
The initial loss for the DNN in Fig.~\ref{fig:ML} is higher than that for the RNN, but this is a function of architecture - the initial loss typically becomes comparable to LSTM/GRU initial losses as the number of layers varies from 1 up to 10.
It is interesting to note that the cascade decay structure is crucial to the improved performance of the RNN setup. For instance, we can consider the separation of $t\bar t Z$ from $t\bar t h$ production. For our NN input data the discrimination is driven by the invariant $b\bar b$ mass, while both processes have a comparable resonance structure. In this case, a comparison of RNN and DNN architectures does not single out the RNN as a better adapted approach.

A strong test of our setup is its capability to isolate the resonance structures from the provided input data in the presence of the significant degradation from missing energy. 
We define a reconstructed $M_{H_3}$ by adding the four-momenta of the two leptons and the four b-tagged jets of highest $p_T$ as well as the missing transverse momentum. Similarly, $M_{H_2}$ is defined from two leptons and a pair of b-tagged jets incompatible with the 125 GeV Higgs mass, $125 \pm 10$~GeV. We can use these definitions to check that we indeed get a peaklike structures after training and selection. The reconstructed masses are shown in Fig.~\ref{fig:reso_histo}, before and after the application of an LSTM network to select events. Although the resonance structure becomes significantly distorted due the sizeable missing energy that arises from a range of sources, the resonance peaks are visible, and the backgrounds are significantly reduced in a signal-like selection.

To determine the sensitivity quantitatively, we perform an analysis of signal and background rates after the application of the LSTM. The cross sections obtained after the LSTM selection 
which is chosen to maintain a large $\sigma_S / \sigma_B$ ($\sigma_S$ and $\sigma_B$ are the signal and background cross sections after selection, respectively). This is done to minimise impact of background systematics which we neglect in this study. Subsequently, we perform a pseudomeasurement by evaluating the signal (background) number of events $S$ ($B$) at an extrapolated integrated luminosity of $3$/ab and determining the significance $S / \sqrt{B}$. For an LSTM network of one layer with 45 units, we obtain a significance of $5.3$ based on a rate of $S/B \simeq 0.09$. Performing the same analysis with a DNN network of two dense layers with 80 units each, the significance is $S / \sqrt{B} = 4.1$ at $S/B \simeq 0.08$. This shows that the DNN is slightly more vulnerable to background systematics, while the GRU/LSTM architecture is essential to claim a new physics discovery in this channel at the HL-LHC. Finally, for comparison, we additionally perform a simple cut-and-count analysis to conclude our comparison of different approaches. Besides the selection criteria, additional cuts are imposed on the missing energy requiring $\slashed{E}_T > 30$~GeV. The search region is further constrained by applying cuts on the transverse momentum of final state particles. The four b-jets must satisfy staggered cuts $p_T(b_1) > 100$~GeV, $p_T(b_2) > 70$~GeV, $p_T(b_3) > 65$~GeV, and $p_T(b_4) > 50$~GeV,  while for the leptons, similarly $p_T(\ell_1) > 30$~GeV and $p_T(\ell_2) > 10$~GeV were imposed. A Higgs compatible pair is reconstructed by requiring the invariant mass of a pair of b-jets to be within $125 \pm 10$~GeV. If more than one possible pair is identified, the one with the smallest separation $\Delta R$ is selected and if no candidate pair is found the event is vetoed. The reconstructed Higgs must satisfy $p_T(h) > 120$~GeV and the invariant mass of the remaining two b-jets is restricted to $m_{bb} > 80$~GeV. The aforementioned cuts result in a smaller $S/B$ ratio compared to the network approaches, evaluated as $0.04$ which corresponds to a significance of $2.1$. This poorer performance highlights the relevance of using as much information as possible in discriminating signal from background as given by the LSTM/GRU and DNN networks to gain sensitivity to new physics scenarios such as the N2HDM at the LHC.

\section{Discussion and Conclusions}
\label{sec:conc}
The search for new physics beyond the Standard Model in scenarios with exotic scalars in cascade decays
 can be subject to interference effects in the best motivated top final states if additional scalars
are heavy. In this case, multi-Higgs production can come to the rescue, as it will provide additional sensitivity to the `standard' SM-like Higgs searches. Furthermore, gaining sensitivity to such decays is crucial for the reconstruction of the underlying microscopic theory. Particularly motivated in this context are decays of a heavy Higgs state into a pair of different mass Higgs bosons, one of which is the 125~GeV state. Such signatures probe particular aspects of the models' UV structure such as 2HDM alignment or an extension of the 2HDM scalar sector, thus also helping to discriminate between different model hypotheses if a discovery is made.

In this work, we have exploited the memory imprinted by cascade decay patterns of a heavy state through a chain of 
decay steps into SM matter. We have demonstrated that RNNs which access this memory in a particularly adapted way exhibit superior discriminative power than `ordinary' DNNs, which would need to learn the decay steps indirectly through 
correctly pairing final state objects. In general, this results in a slightly reduced sensitivity of DNN networks for the considered physics case. In parallel, the DNN performance results from a longer learning period while the RNNs pick up the available information rapidly. To highlight the physical relevance of this approach we have considered a parameter point of the N2HDM model space that could be observed in the cascade decay channel using the RNN approach with a significance of over $5\sigma$ at the HL-LHC. The RNN architecture is particularly relevant for this parameter point to be able to claim a discovery at the LHC. While we have focused on a particular decay chain, our results can be expected to generalise to the other UV scenarios such as the NMSSM where the scalar mass scales are different, and $H_2$ would lie below the 125 GeV boson with direct decays $H_2\to b\bar b$. We leave this for future work.

\bigskip
\noindent{\bf{Acknowledgments}} ---
C.E. is supported by the UK Science and Technology Facilities Council (STFC) under grant ST/P000746/1 and by the IPPP Associateship Scheme.
M.S. is supported by the STFC under grant ST/P001246/1. 
P.S. is supported by an STFC studentship under grant ST/T506102/1.
The work of M.F. is funded partly by the STFC GrantST/L000326/1. 
S.V. and M.F. are also supported by the European Research Council under the European Union’s Horizon 2020 programme (ERC Grant Agreement no. 648680 DARK-HORIZONS). 
S.V. was the recipient of a Sir Richard Trainor Scholarship at the start of her PhD.

\bibliography{references} 

\end{document}